\documentclass[prd,preprint,tightenlines,showpacs]{revtex4}
\RequirePackage{amsfonts}
\RequirePackage{amssymb}
\RequirePackage{amsmath}
\RequirePackage{graphicx}
\RequirePackage{makeidx}
\RequirePackage{ifthen}
\RequirePackage{color}
\renewcommand{\a}{\ensuremath{\alpha}}
\renewcommand{\b}{\ensuremath{\beta} }  
\newcommand{\g}{\ensuremath{\gamma}  }
\renewcommand{\d}{\ensuremath{\delta}  }

\renewcommand{\th}{\ensuremath{\theta} }

\newcommand{\G}{\ensuremath{\Gamma}  } 
\newcommand{\D}{\ensuremath{\Delta}   }

\newcommand{\be}{\begin{eqnarray}}          \newcommand{\ee}{\end{eqnarray}}
\newcommand{\ba}{\begin{align}}          \newcommand{\ea}{\end{align}}
\newcommand{\benum}{\begin{enumerate}}   
\newcommand{\eenum}{\end{enumerate}}
\newcommand{\bitem}{\begin{itemize}}          
\newcommand{\eitem}{\end{itemize}}

\newcommand{\nn}{\nonumber \\ }

\newcommand{\abs}[1]{\ensuremath{\left| #1 \right|}}   

\newcommand{\sci}[2]{\ensuremath{#1\times 10^{#2}}}  
 
\newcommand{\order}[1]{\ensuremath{{\cal O}(10^{#1})} }

\newcommand{\fr}[2]{\ensuremath{\frac{#1}{#2}}}

\newcommand{\inv}[1]{\frac{1}{#1}}        

\renewcommand{\matrix}[1]{\begin{pmatrix} #1 \end{pmatrix}}

\definecolor{myblue}{rgb}{.1,.1,.7}
\definecolor{dcyan}{rgb}{.0,.6,.6}
\definecolor{dmagenta}{rgb}{0.6,0.0,0.6}
\definecolor{brown}{rgb}{0.6,0.2,0.}
\definecolor{darkblue}{rgb}{0,0,0.6}
\definecolor{darkred}{rgb}{0.75,0.0,0.0}
\definecolor{darkgreen}{rgb}{0.0,0.6,0.0}

\newcommand{\blue}{\color{blue}}

\newcommand{\usemarker}{Y}
\newcommand{\marker}[1]{
       \ifthenelse{\equal{\usemarker}{Y}}
                     {\mbox{}\marginpar{\tt #1}}{}
               }

\newcommand{\mk}[1]{
\ifthenelse{\equal{\usemarker}{Y}}
{\noindent\hskip -1truecm {\bf\blue$^{#1}$}}{}
}

\newcommand{\myname}{Yu-Feng Zhou}

\newcommand{\myaddress}{Ludwig-Maximilians-University Munich, \\
Sektion Physik. Theresienstr. 37, D-80333. Munich, Germany}

\renewcommand{\usemarker}{N}
\reversemarginpar
\usepackage{feynmf}
\begin{document}\begin{fmffile}{fmftemp}
\title{Texture of Yukawa coupling matrices  in  general  two-Higgs-doublet model}
\author{\myname}
\affiliation{\myaddress}
\date{\today}  
\begin{abstract} 
  We discuss possible parallel textures of the Yukawa coupling matrices in the general
  two-Higgs-doublet model (2HDM).  In those textures the flavor changing neutral currents
  are  naturally suppressed.  Motivated by a phenomenologically successful texture
  with four texture zeros  in the standard model, 
 we propose a predictive ansatz for the Yukawa coupling matrices with the same texture in the 
general 2HDM.  Compared with the six texture-zero
  based ansatz proposed by Cheng and Sher, it is in a better agreement with the data of
  quark mixings and CP violation.  The four texture-zero based ansatz predicts a  different
  hierarchy in the Yukawa coupling matrix elements.  As a consequence, in the lepton sector, the
  related Yukawa couplings are less constrained by the experimental upper bound of $\mu\to e\gamma$,
  which allows  significantly larger predictions for other processes.   The contributions
  from neutral scalar interactions to the lepton number violation decay modes $\ell\to \ell_{1}\ell_{2}\ell_{3}$ are
  calculated in both ansatz.   It is shown  that the predictions from  the four texture-zero based ansatz
   could be  two order of magnitude greater than that from the six texture-zero
  based one.  The branching ratio of $\mu\to 3e$ and $\tau\to 3\mu$ can reach \sci{7.5}{-17} and \sci{1.3}{-10} respectively.
The predicted ratio of $Br(\mu\to 3e)/Br(\tau\to 3e)$ is also
  larger and almost parameter independent.  Those differences make the two ansatz to be
  easily distinguished by the future experiments.
\end{abstract}
\preprint{LMU 03-13}
\maketitle
  
\section{introduction} 
  
Although the standard model (SM) of electroweak interactions 
with $SU(2)_{L}\otimes U(1)_{Y}$ guage symmetry
has achieved a great success in
phenomenology, the outstanding problems such as the origin of the fermion masses,
the mixing angles as well as the CP violation are still unresolved. It is widely believed that
the SM can not be a fundamental theory of the basic interactions.  For this reason, many
new physics models have been proposed and extensively studied in the recent years.  Among
those models, the SM with two-Higgs-doublet (2HDM) can be regard as  the simplest extension
of the present  SM, and have attracted a lot of   attention (see,  e.g.
\cite{
Glashow:1977nt,
cheng:1987rs,Antaramian:1992ya,Hall:1993ca,wu:1994ja,
atwood:1997vj,wu:1999fe,Bowser-Chao:1998yp,%
Hou:1992un,Atwood:1996ud,Atwood:1995ej,%
wolfenstein:1994jw,Atwood:1996vw,%
Dai:1997vg,Iltan:2000gy,Bobeth:2001sq,Zhou:2000ym,%
Diaz:2001qb,Cheung:2001hz,Wu:2001vq, Zhou:2001ew,%
Sher:2000uq,Larios:2002ha,Xiao:2002mr}
). 

In the most general form of 2HDM \cite{cheng:1987rs,Antaramian:1992ya,Hall:1993ca,wu:1994ja}, the Lagrangian for the  Yukawa interaction is given by 
\begin{align}\label{2HDMLagrangian}
-\mathcal{L}_{Y}&=
\bar{q}_{_L} \G^{u}_{1} \widetilde{\phi}_{1} u_{_R}
+ \bar{q}_{_L} \G^{d}_{1} \phi_{1} d_{_R}
+ \bar{q}_{_L} \G^{u}_{2} \widetilde{\phi}_{2} u_{_R}
+ \bar{q}_{_L} \G^{d}_{2} \phi_{2} d_{_R}
\nn
&+ \bar{\ell}_{_L} \G^{e}_{1} \phi_{1} l_{_R} 
+ \bar{\ell}_{_L} \G^{e}_{2} \phi_{2} l_{_R}
+\mbox{H.c},
\end{align}
with $\G^{F}_{i} (i=1,2)$ being the Yukawa interaction matrices and  
$\widetilde{\phi}_{i}=i \tau_{2} \phi^{*}_{i}$.
The two Higgs fields after the spontaneous symmetry breaking have the following form
\begin{align}
\phi_{1}=\matrix{\phi^{+}_{1} \\ \fr{1}{\sqrt2}( v_{1}e^{i\d}+\phi^{0}_{1}+i \chi^{0}_{1})},
\qquad
\phi_{2}=\matrix{\phi^{+}_{2} \\ \fr{1}{\sqrt2}( v_{2}+\phi^{0}_{2}+i \chi^{0}_{2})},
\end{align}
where $v_{1}$ and $v_{2}$ are the absolute values of the vacuum expectation values (VEVs)
of the two Higgs fields, which satisfy $v=\sqrt{v_{1}^2+v_{2}^{2}}\simeq246$ GeV.  $\d$ is a
relative phase between them, which can be a new source of CP violation in this model\cite{wu:1994ja}.
The ratio between $v_{1}$ and $v_{2}$ is often referred to as $\tan\b$ with the definition
of $\tan\b\equiv v_{2}/v_{1}$.

In the general 2HDM, the  relation to the Higgs mechanism  in the SM is manifest if    
one  applies a  basis transformation of
\begin{align}
H=\phi_{1}e^{-i\d} \cos\b  + \phi_{2} \sin\b
, \quad \mbox{and} \quad
\phi=\phi_{1}e^{-i\d} \sin\b - \phi_{2}\cos\b.
\end{align}
In this  new basis, the two Higgs fields are
\begin{align}
H=\matrix{G^{+} \\ \\ \inv{\sqrt2}(\rho^{0} +i G^{0} + v )}
, \quad
\phi=\matrix{H^{+} \\ \\  \inv{\sqrt2}( R^{0}+ i I^{0})},
\end{align}
where $ G^{\pm}, G^{0}$ are the would-be  Goldenston bosons in 
the SM. $H^{+}$, $R^{0}$ and $I^{0}$ are new scalars.
In general, the flavor eigenstates $\hat H=(\rho^{0}, R^{0}, I^{0})$ are not the mass
eigenstates. The mass eigenstates  denoted by $H=(H^{0}, h^{0}, A^{0})$ are related to 
the flavor eigenstates through a mixing matrix $O$, i.e.  $H_{i}=O_{ij}\hat H_{j}$. 

In the whole discussion below, for the sake of simplicity, we will focus on the case in
which  both VEVs are real, i.e. $\d=0$, and the effects of neutral scalar mixing are
negligible.  To simplify notations, we omit the flavor index of $F=u, d, e$ in the
superscript of the Yukawa matrices $\G^{F}_{i}$ in Eq.(\ref{2HDMLagrangian}), and will add
them latter when a distinction between flavors is needed. 

In this approximation, the fermion mass and the Yukawa coupling  matrices are simply given by
\begin{align}\label{massInSM}
M&=\fr{v}{\sqrt2}(\G_{1}\cos\b+\G_{2}\sin\b),
\nn 
\G&=\fr{1}{\sqrt{2}}(\G_{1}\sin\b- \G_{2} \cos\b).
\end{align}

The main problem in  the models with multi-Higgs-doublet  is that in general
$M$ and $\G$ can not be diagonalized simultaneously,  the off diagonal elements
of $\G$ after being rotated into mass basis  may give too large contributions to $K^{0}-\bar K^{0}$ mixing and
$K_{L}\to \mu\mu$ via flavor changing neutral currents (FCNCs)   at tree level.
To forbidden the tree level FCNC in 2HDM,  the  {\it ad hoc} discrete symmetries are 
often imposed on the Lagrangian 
\cite{Glashow:1977nt}
\begin{align}\label{model12}
 \phi_1 \rightarrow -\phi_1 & \mbox{\ and \ }  \phi_2 \rightarrow \phi_2,
\nn
 u_{_{R}} \rightarrow -u_{_{R}} & \mbox{\ and \ } d_{_{R}} \rightarrow \mp d_{_{R}},
\end{align}
which defines two types of 2HDM without tree level FCNCs,   referred to as model I and II
of 2HDM.  However, since at present the most strict bound of tree level FCNCs only comes from the
light quark sector, There is a possibility that the tree level FCNCs do occur, but
greatly suppressed due to  the smallness of the light quark masses. Abandoning the discrete
symmetry in Eq.(\ref{model12}), one arrives at the general type of 2HDM with small FCNCs
\cite{cheng:1987rs,Antaramian:1992ya,Hall:1993ca,wu:1994ja}.
The small off diagonal Yukawa matrix elements can be attributed to an approximate flavor
symmetry which is slightly broken down, and the magnitudes of the symmetry breaking are
proportional to the related fermion masses \cite{Antaramian:1992ya,Hall:1993ca,wu:1994ja}.

Besides the symmetric considerations, another  practical way  to prevent from
large FCNCs  in the general 2HDM is to adopt reasonable anastz on the textures of the Yukawa interaction
matrices $\G_{i}$.  It is well known that without a loss of generality,  the mass
matrices $M$ can be rotated to be Hermitian by a  suitable redefinition of the fermion
fields in the flavor basis \cite{Frampton:1985qk}. In the case of both  VEVs are real and arbitrary ,
$\G_{1}$ and $\G_{2}$ could  also be rotated to be Hermitian.  The possibility that
$M$ and $\G$ be diagonalized simultaneously depends only on the commutator of them,
namely they can be diagonalized simultaneously if and only if $[M,\G]=0$. From Eq.(\ref{massInSM}), the
commutator of $M$ and $\G$ is given by
\begin{align}
[M, \G]=-\fr{v}{2}[\G_{1}, \G_{2}].
\end{align}
Obviously, if $\G_{i}$s  have a parallel structure  of
\begin{align}
\G_{2}\approx c
\G_{1}, 
\end{align}
with $c$ being a number factor, the commutator will  be close to zero,  the Yukawa
coupling matrix $\G$ after being rotated into mass basis
will be nearly diagonal, and the tree level FCNCs  will be suppressed accordingly.  The
simplest way to get the  parallel texture is to impose a $S_{2}$ permutation symmetry between
$\phi_{1}$ and $\phi_{2}$ on the Lagrangian,  namely  the Lagrangian is invariant under the 
transformation  of
\begin{align}
\phi_{1} \longleftrightarrow \phi_{2},
\end{align}
which directly leads to $c=1$, and $\G_{1}=\G_{2}$. Thus all the tree level FCNCs vanish.
 
To get small off diagonal elements, such an symmetry must be slightly broken down.
The breaking down of the exact parallel texture can be parameterized as follows
\begin{align}\label{parallel}
\G_{2}=c \G_{1}- \D \G,  \qquad \mbox{ with } (\D\G)_{ij}/(\G_{1})_{ij} \ll 1,
\end{align}
where $\D \G$ is a small perturbation matrix which does not commutate with $\G_{1}$. In 
this parameterization, the commutator is given by
\begin{align}
[M, \G]=\fr{v}{2}[\G_{1}, \D\G] \neq 0,
\end{align}
and the small tree level FCNCs could appear. 
The Hermitian mass matrix $M$ is diagonalized by a rotation matrix $R$  such that
\begin{align}
R^{\dagger} M R= 
\matrix{ 
m_{1} & 0 & 0 \\
0 & m_{2}& 0 \\
0 & 0 & m_{3} 
}.
\end{align}
where $m_{1}, m_{2}$ and $m_{3}$ are the eigenstates  which are equal to the 
three fermion masses  up to a sign difference.
Applying the same transformation to $\G$, one
arrives at the Yukawa matrix in the mass basis
\begin{align} 
Y\equiv & R^{\dagger} \G R 
\nn
=& \inv{v} 
\left(
\fr{\sin\b-c\cos\b}{\cos\b+c\sin\b}
\right)
\matrix{
m_{1} & 0  & 0 \\ 
0  &  m_{2}  & 0 \\ 
0  & 0  & m_{3} 
}
+\fr{R^{\dagger} \D\G R}{\sqrt2(\cos\b+c\sin\b)} .
\end{align}
Thus from the approximate parallel texture of Eq.(\ref{parallel}) the dominant part of the
Yukawa coupling matrix in mass basis is a diagonal one, and the off diagonal elements
are determined by the small matrix $\D \G$. Furthermore,  the diagonal part of
the Yukawa coupling matrix has the same hierarchical structure as the one in the fermion
mass matrix,  namely the $(i, i)$ element is proportional to $m_{i}$.
 
Another advantage of taking the parallel texture of Yukawa coupling matrices is that the mass
matrix $M$ and the Yukawa matrix $\G$ will be closely related.  Given a phenomenologically
successful texture of the mass matrix, the form of matrix $\D \G$ can be inferred, and
 the  off diagonal elements can be obtained as well.
 
There are already extensive studies on the possible textures of the mass matrices in the
SM. The most simplest ones among them are the textures with texture-zeros 
( see, e.g. 
\cite{
Fritzsch:1977za,Fritzsch:1978vd,Fritzsch:1979zq,Ramond:1993kv,%
Gupta:1991cq,Giudice:1992an,Fritzsch:1999rb,Rosenfeld:2001sc,Roberts:2001zy}
).  
It is well
known that a simple zero in the (1,1) element of the quark mass matrices leads to a
correct prediction of the Cabibbo angle of $\th_{C}\simeq \sqrt{m_{d}/m_{s}}$ in the two family case
 \cite{Fritzsch:1977za}.  In the three family case, a similar texture with six texture-zeros is widely discussed which is often referred to as "Fritzsch matrix" \cite{Fritzsch:1978vd,Fritzsch:1979zq}. It takes the
following form
\begin{align}\label{6zeroSM}
M=\matrix{
0         &  D     &  0  \\
D^{*}  &  0        & B  \\
0         & B^{*}  &  A 
},
\end{align}
where $A\simeq m_{3}$, $B\simeq \sqrt{m_{2}m_{3}}$
and $D \simeq \sqrt{m_{1}m_{2}}$.   
 
Motivated by this texture in the SM, an ansatz of the Yukawa matrix is proposed by Cheng
and Sher \cite{cheng:1987rs}, which suggests a similar structure of the both Yukawa matrix
in the flavor basis
\begin{align}\label{6zero2HDM}
\G_{i}=\fr{1}{v_{i}}\matrix{
0         & d_{i} D   &  0  \\
d^{*}_{i}D^{*}    &  0        & b_{i}B \\
0         & b^{*}_{i}B^{*} & a_{i} A
}
\qquad (i=1,2),
\end{align}
 
%
The free parameters $a_{i},b_{i}$ and $d_{i}$ are assumed to be at the same order
of magnitudes of order one.  In order to reproduce the quark mass matrix,
they  must satisfy the relation of $\sum a_{i}=\sum b_{i}=\sum d_{i}=1$.   
Clearly, the two Yukawa matrices $\G_{1}$ and $\G_{2}$ have a  parallel texture.
After being rotated into the mass basis, the Yukawa 
matrix  takes the form of 
\begin{align}\label{6zero}
Y\simeq \fr{1}{v} 
\matrix{
  \xi_{_{11}} m_{1}  & \xi_{_{12}} \sqrt{m_{1} m_{2}}   &  \xi_{_{13}} \sqrt{m_{1}m_{3}} \\
  \xi^{*}_{_{12}}\sqrt{m_{1}m_{2}}   & \xi_{_{22}} m_{2} &  \xi_{_{23}} \sqrt{m_{2}m_{3}}\\
   \xi^{*}_{_{13}}\sqrt{m_{1}m_{3}}   & \xi^{*}_{_{23}}\sqrt{m_{2}m_{3}}& \xi_{_{33}} m_{3}  
},
\end{align}
%
with $\xi_{ij}$s being the functions of $a_{i}, b_{i}$ and $d_{i}$, and 
have the following values \cite{cheng:1987rs}
\begin{align}
\xi_{11}&=(2d_{1}-2b_{1}+a_{1})\tan\b
-(2d_{2}-2b_{2}+a_{2})\cot\b,
\nn
\xi_{12}&=(d_{1}-2b_{1}+a_{1})\tan\b
-(d_{2}-2b_{2}+a_{2})\cot\b,
\nn
\xi_{13}&=\xi_{23}=(b_{1}-a_{1})\tan\b 
-(b_{2}-a_{2})\cot\b,
\nn
\xi_{22}&=(a_{1}-2b_{1})\tan\b
-(a_{2}-2b_{2})\cot\b,
\nn
\xi_{33}&=a_{1}\tan\b -a_{2}\cot\b.
\end{align}
If no accidental cancellations, all of  the $\xi_{ij}$s are roughly  in  the same order of magnitude.
To clearly see the hierarchy in the Yukawa matrix,  
it is  convenient to introduce a common factor  $\xi$ with
\begin{align}
\xi_{ij}\approx \xi  \qquad (i,j=1,2,3).
\end{align}
Thus the Yukawa coupling matrix can be approximately  written as
\begin{align}\label{6zeroSimple}
Y\approx \fr{\xi}{v} 
\matrix{
  m_{1}  &  \sqrt{m_{1} m_{2}}   &  \sqrt{m_{1}m_{3}} \\
  \sqrt{m_{1}m_{2}}   &  m_{2} &    \sqrt{m_{2}m_{3}}\\
   \sqrt{m_{1}m_{3}}   & \sqrt{m_{2}m_{3}}&   m_{3}  
},
\end{align}
%
%
%
One finds that the hierarchical structure of $Y$ is dominated by the fermion masses.  The
FCNCs related to light fermions are greatly suppressed by the small fermion masses. For
example, for $d \bar s h^{0}$ transition, the Yukawa coupling is of the order of
$\order{-6}$, which successfully explains the almost invisible  FCNC in neutral kaon
mixing \cite{cheng:1987rs,atwood:1997vj,wu:1999fe}.  But for $t \bar c h^{0}$ transition,
it could reach order one, and predicts  possible large FCNCs in the related processes
\cite{Hou:1992un,Luke:1993cy,Atwood:1995ej,Atwood:1996ud}.

The above ansatz appears to be simple, symmetric and attractive.  However, the latter
studies have already shown that the base of this ansatz, i.e. the six texture-zero texture
is in severe problem in accounting for the current data of the CKM matrix elements from
the  known quark masses ( for detailed discussion see, e.g \cite{Harari:1987ex}).
The texture of Eq.(\ref{6zeroSM}) leads to the following predictions of the CKM matrix
elements \cite{Fritzsch:1979zq}
\begin{align}\label{6-zero-prediction}
\abs{V_{ud}} &\simeq \abs{\sqrt{\fr{m_{u}}{m_{c}}}-e^{-i \phi_{1}}\sqrt{\fr{m_{d}}{m_{s}}}},
\nn 
\abs{\fr{V_{ub}}{V_{cb}}}&\simeq \sqrt{\fr{m_{u}}{m_{c}}},
\qquad 
\abs{\fr{V_{td}}{V_{ts}}}\simeq \sqrt{\fr{m_{d}}{m_{s}}},
\nn
\abs{V_{cb}}&\simeq \abs{\sqrt{\fr{m_{s}}{m_{b}}}-e^{-i\phi_{2}}\sqrt{\fr{m_{c}}{m_{t}}}}.
\end{align}
where $\phi_{1}$ and $\phi_{2}$ are two phase parameters.
Taking the values of the quark masses  from Ref.\cite{Hagiwara:2002fs}, it is found that the first prediction is in a very
good agreement with the experimental data.  The next two  relations roughly agree with the data as well, but the
last one contradicts to  the current experiment. It has already being pointed out that for a large 
value of $m_{t} \gtrsim 90$ GeV,  it is impossible to  fine-tune the values of 
the quark masses to achieve  an agreement \cite{Harari:1987ex}. For a large value of $m_{t}=175$ GeV, the
six texture-zero based mass matrix gives $\abs{V_{cb}} \approx \sqrt{m_{s}/m_{b}}\approx 0.2$
which is too large compared  with the current data of $V_{cb}= 0.04\pm0.002$.

The other problem of the six texture-zero texture is that there is no room for  CP violation in the
Yukawa sector\cite{Fritzsch:1979zq}.  This is because the Jarlskog rephasing invariant
quantity of CP violation \cite{Jarlskog:1985ht} $\mathcal{J}_{CP}$ is proportional to $
\mbox{det}[M^{u},M^{d}]$ which  is always zero in this texture.  All the complex phases in the
mass matrix elements are unphysical, and can be rotated away in the general 2HDM.  If the six texture-zero texture
is adopted in the Yukawa matrices $\G_{1,2}$,  there will still be no
CP violation in the Yukawa sector even  when the relative phase $\d$ between the two VEVs is nonzero. For these
reasons, this texture has been completely ruled out.

In order to accommodate all the current data of quark masses and mixing angles in the
framework of texture-zeros, a natural choice is to give up the texture zero in (2,2)
element of the mass matrices. Keeping the texture of up and down fermion mass matrices to be parallel, one
arrives at the so called four texture-zero textures which can explain all the current CKM
matrix elements within  a reasonable precision \cite{Froggatt:1979nt,Gill:1995pn,Fritzsch:1995nx,Fritzsch:1999rb,Chiu:2000gw}.  This texture is given by
\begin{align}\label{4zeroSM}
{M}=\matrix{
0         &   D   &  0  \\
 D^{*} &   C   &  B' \\
0         &   B'^{*}   &   A
} ,
\end{align}
with $| B'|\simeq | C|\simeq m_{2}$. 
This kind of textures can also be obtained from  symmetric considerations
such as $U(2)$ horizontal symmetry \cite{Pomarol:1996xc,Caravaglios:2002br}
or other symmetries \cite{Chou:1996di}.
In this four texture-zero texture, the
predicted value of $|V_{cb}|$ is $|V_{cb}|\simeq m_{s}/m_{b}\simeq 0.04$ in a good agreement
with the experiment. Furthermore, in this texture there exists nontrivial complex phases which
can directly result in $\mathcal{J}_{CP}\neq 0$. Thus CP violation can be accommodated. 

Using  the four texture-zero texture instead of the six texture-zero based one, the 
Eq.(\ref{6zero2HDM}) is rewritten as
\begin{align}\label{4zero2HDM}
{\G}_{i}=\inv{v_{i}}\matrix{
0         & d_{i}   D   &  0  \\
 d^{*}_{i}  D^{*} &   c_{i} C   &  b_{i} B' \\
0         &  b^{*}_{i}  B'^{*}   &   a_{i} A
}.
\end{align}
Following the similar procedures from Eq.(\ref{6zeroSM}) to (\ref{6zero}), 
one finds that the corresponding Yukawa coupling matrix
in the mass basis is given by
\begin{align}\label{4zero}
 Y' \simeq \fr{1}{v} 
\matrix{
  \xi'_{_{11}}m_{1}  & \xi'_{_{12}} \sqrt{m_{1} m_{2}}   &  \xi'_{_{13}} \sqrt{m_{1}m_{2}} \\
  \xi'^{*}_{_{12}}\sqrt{m_{1}m_{2}}   & \xi'_{_{22}} m_{2} &  \xi'_{_{23}}m_{2} \\
  \xi'^{*}_{_{13}} \sqrt{m_{1}m_{2}}   & \xi'^{*}_{_{23}}m_{2} & \xi'_{_{33}} m_{3}  
} ,
\end{align}
with 
\begin{align}
\xi'_{11}&=(c_{1}-2d_{1})\tan\b
-( c_{2}-2d_{2})\cot\b,
\nn
\xi'_{12}&=(-c_{1}+d_{1})\tan\b
-(-c_{2}+d_{2} )\cot\b,
\nn
\xi'_{13}&=(a_{1}-b_{1})\tan\b
-(a_{2}-b_{2})\cot\b,
\nn
\xi'_{22}&=c_{1}\tan\b
- c_{2}\cot\b,
\nn
\xi'_{23}&=(-a_{1}+b_{1})\tan\b
-( -a_{2}+b_{2})\cot\b,
\nn
\xi'_{33}&=a_{1}\tan\b -a_{2}\cot\b.
\end{align}
Similarly, all the couplings $\xi'_{ij}$ are roughly at the same order of magnitude,
and  can be approximately written as 
\begin{align}
\xi'_{ij}\approx  \xi'.
\end{align}
The Yukawa coupling matrix is then simplified to be  
\begin{align}\label{4zeroSimple}
 Y'\approx \fr{\xi'}{v} 
\matrix{
  m_{1}  &  \sqrt{m_{1} m_{2}}   &   \sqrt{m_{1}m_{2}} \\
   \sqrt{m_{1}m_{2}}   &  m_{2} &  m_{2} \\
   \sqrt{m_{1}m_{2}}   &  m_{2} &  m_{3} 
}.
\end{align}


Comparing Eq.(\ref{4zeroSimple}) with Eq.(\ref{6zeroSimple}), one finds that
although both ansatz lead to the same structure of the Yukawa couplings to the first and
the second generation fermions, the couplings to the third generation fermions in
Eq.(\ref{4zeroSimple}) are relatively much weaker.  The relative suppression of the Yukawa
couplings to the third generation fermion in Eq.(\ref{4zeroSimple}) can be understood in
three aspects:
Firstly, 
in the four texture-zero based ansatz,   the ratio of the couplings between  (1,1), (1,2) and (1,3) 
element is  
$m_{1}:\sqrt{m_{1}m_{2}}: \sqrt{m_{1}m_{2}}$ 
while in Cheng-Sher ansatz it is
$m_{1}: \sqrt{m_{1}m_{2}}:\sqrt{m_{2}m_{3}}$. 
Thus compared with the one in Cheng-Sher
ansatz, the (1,3) elements is smaller by a factor of $\sqrt{m_{3}/m_{2}}$.
Secondly,  The four texture-zero based ansatz leads the ratio between (2,2) and (2,3) element to be $1:1$ 
, but  the ratio in Cheng-Sher ansatz is $\sqrt{m_{2}}: \sqrt{m_{3}}$. 
The (2,3) elements is also suppressed  by a the same factor of $\sqrt{m_{3}/m_{2}}$.
Finally, in Eq.(\ref{4zeroSimple}) both the (1,3) and (2,3) elements are relatively
smaller compared with its (3,3) element. For the ratio between (1,3), (2,3) and (3,3)
element, the four texture-zero based ansatz gives $\sqrt{m_{1}m_{2}}: m_{2}:m_{3}$ but the
one in Cheng-Sher ansatz is $\sqrt{m_{1}m_{3}}:\sqrt{m_{2}m_{3}}:m_{3}$.
Those differences will lead to significantly different consequences in phenomenology.

Note that the complete  Yukawa coupling matrix contains  a
global factor of $\xi(\xi')$ which has  to be constrained or determined from the
experiments. 
In the quark sector the most strict constraint comes from the light quark sector,
especially $K^{0}-\bar K^{0}$ mixing ( for detailed discussions, see,
e.g.\cite{wu:1999fe}). The resultant bounds on the Yukawa coupling for down quarks $\xi^{d}$
and $\xi^{d'}$ are the same, as both ansatz have the same hierarchy in the couplings to
the first and the second generations.  Therefore, due to the relative suppression of the
$(1,3)$ and $(2,3)$ elements in the Yukawa coupling matrices, the predictions from the
four texture based ansatz in Eq.(\ref{4zeroSimple}) for processes involving heavy fermions
such as rare $B$ meson and $t$ quark decays will be much smaller.

\begin{figure}[htb]
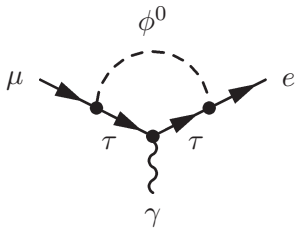

\begin{fmfchar*}(30,30)
  \fmfleft{v1}\fmflabel{$\mu$}{v1}
  \fmfright{v2}\fmflabel{$e$}{v2}
  \fmfbottom{v3}\fmflabel{$\gamma$}{v3}  
  \fmf{fermion}{v1,o1}\fmf{fermion,label=$\tau$}{o1,z}
  \fmf{fermion,label=$\tau$,label.side=right}{z,o2}\fmf{fermion}{o2,v2}
  \fmf{dashes,left,tension=0,label=$\phi^{0}$}{o1,o2}
  \fmf{photon}{z,v3}
  \fmfdot{o1,o2,z}
\end{fmfchar*}
\caption{Feynman diagram for netral scalar contribution to $\mu\to e \g$.}
\label{mueg}
\end{figure}

On the other hand, in the lepton sector, the strongest constraint comes  from the radiative
decay $\mu\to e\gamma$,  which is relevant to the (1,3) and (2,3) elements of the Yukawa
coupling matrix  $Y$. In this case, the value of 
the Yukawa couplings for leptons $\xi^{e'}$ will be less constrained and
the allowed range will be greater than that of $\xi^{e}$ by a factor of $\sqrt{m_{\tau}/m_{\mu}}$. As a result, 
the predictions to the precesses  relating  to the first and the second generation fermions could 
be {\it enhanced} rather than suppressed.  The predicted branching ratios of those processes 
will be  greatly
enlarged as they depend on the fourth power of $\xi^{e}(\xi^{e'})$, which may lead to significantly 
different predictions, and make the texture in Eq.(\ref{4zeroSimple}) and Eq.(\ref{6zeroSimple}) to 
be easily distinguished   by the future experiments.
 %
The Feynman diagram for the general 2HDM contribution to $\mu\to e \gamma$ at one loop
is shown in Fig.\ref{mueg}.
In the general 2HDM,  the tree level FCNC presents, which allows the $\mu$ lepton to go  into
much heavier $\tau$ lepton  at the one loop level, and gives a large contribution from   loop
integration. Thus in the general 2HDM,  the one loop flavor changing diagrams  dominates.
Note that it is quite  different from the case in the 2HDM of type II, in which the two
loop Barr-Zee diagrams \cite{Bjorken:1977vt,Barr:1990vd} give the dominant contributions.
This is because in 2HDM of type II the tree level FCNC is forbidden, at one loop level ,the neutral scalars
must couple to light leptons, and the Yukawa couplings which are proportional
to the  lepton masses are very small.  But at two loop level the scalars may couple
to heavy quark or leptons in a second fermion loop, in which large Yukawa couplings may
present \cite{Zhou:2001ew,Chang:1993kw}.

The branching ratio of $\mu\to e \g$ at one loop level   is given by \cite{Zhou:2001ew}
\begin{align}
Br(\mu\to e\g)=\fr{\a_{em}\tau_{\mu}m_{\tau}^{2}m_{\mu}^{3}}{2^{10}\pi^{4}} \fr{ }{}
\left(\fr{Y_{\mu\tau}Y_{\tau e}}{m_{\phi}^{2}}\right)^{2}\ln^{2}\left(\fr{m_{\phi}^{2}}{m_{\tau}^{2}}-\fr32 \right) 
,
\end{align}
where $\phi=h^{0}$ or $A^{0}$. 
The contributions from the scalar and the pseudo-scalar exchange are the same. Taking the current upper
bound of $Br(\mu\to e\g)<\sci{1.2}{-11}$ \cite{Hagiwara:2002fs}, one finds that
\begin{align}\label{constraint}
\left(\fr{\xi^{e'}}{m_{\phi}}\right) <\sci{6.8}{-3} \mbox{ GeV$^{-1}$},
\qquad
 \left(\fr{\xi^{e}}{m_{\phi}}\right) <\sci{1.7}{-3} \mbox{ GeV$^{-1}$}.
\end{align}
%
Obviously, the allowed value of $\xi^{e'}$ is greater than $\xi^{e}$.   The ratio
of $\xi^{e(')}/m_{\phi}$
obtained here can be directly used to make predictions, since in  all the cases the
predicted decay rates in 2HDM depend only on the ratio rather than the individual values
of the  couplings and the scalar masses.  As   the related processes contain the 
fourth power of $\xi^{e(')}/m_{\phi}$,  the different texture of the Yukawa coupling
matrices may lead to significantly different predictions.

\begin{figure}[htb]
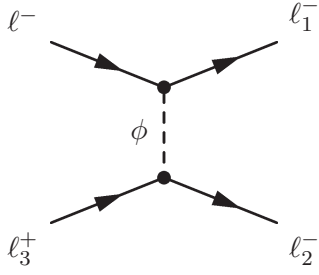
 
\begin{fmfchar*}(30,30)
  \fmftop{L,L1}\fmflabel{$\ell^-$}{L}\fmflabel{$\ell_{1}^{-}$}{L1} 
  \fmfbottom{L3,L2}\fmflabel{$\ell_{3}^+$}{L3}\fmflabel{$\ell_{2}^{-}$}{L2}
  \fmf{fermion}{L,o1,L1}\fmf{fermion}{L3,o2,L2}
  \fmf{dashes,label=$\phi$}{o1,o2}
  \fmfdot{o1,o2}
\end{fmfchar*}
\caption{Feynman diagram for $\ell\to
\ell_{1}\ell_{2}\ell_{3}$ in the general 2HDM.}
\label{3bodyLepton}
\end{figure}

For a concrete illustration, we  take  the 3-body lepton decays $\ell\to
\ell_{1}\ell_{2}\ell_{3}$ as  examples,  and calculate the branching
ratios  in the general 2HDM with two different ansatz in
Eq.(\ref{4zeroSimple}) and (\ref{6zeroSimple}). Those decay modes are forbidden in 
the SM, but in the general 2HDM it could  happen at tree level. Thus
they are very sensitive to the new physics beyond the SM. The related Feynman 
diagram is shown in Fig.\ref{3bodyLepton}.
%
 
The predicted branching ratios  for the various decay modes  are simply given by 
\begin{align}
Br(\mu^{-}\to e^{-}e^{+}e^{-})
&=\fr53 \cdot \fr{\tau_{\mu}}{ 2^{11} \pi^{3}}\fr{m_{\mu}^{5}}{m_{\phi}^{4}}
(Y_{e\mu}Y_{ee})^2,
\nn
Br(\tau^{-}\to e^{-}e^{+}e^{-})
&=\fr53  \cdot \fr{\tau_{\tau}}{ 2^{11} \pi^{3}}\fr{m_{\tau}^{5}}{m_{\phi}^{4}}
 (Y_{e \tau}Y_{ee})^2,
\nn
Br(\tau^{-}\to \mu^{-}e^{+}e^{-})
&=\fr13  \cdot \fr{\tau_{\tau}}{ 2^{10} \pi^{3}}\fr{m_{\tau}^{5}}{m_{\phi}^{4}}
\left[
(Y_{\mu\tau} Y_{ee})^2 
+(Y_{e\tau} Y_{e \mu})^2
+\fr12   Y_{\mu\tau} Y_{ee} Y_{e\tau} Y_{e \mu}
\right],
\nn
Br(\tau^{-}\to \mu^{+}e^{-}e^{-})
&=\fr53  \cdot \fr{\tau_{\tau}}{ 2^{12} \pi^{3}}\fr{m_{\tau}^{5}}{m_{\phi}^{4}}
(Y_{e \tau}Y_{e\mu})^2,
\nn
Br(\tau^{-}\to e^{-}\mu^{+}\mu^{-})
&=\fr13  \cdot  \fr{\tau_{\tau}}{ 2^{10} \pi^{3}}\fr{m_{\tau}^{5}}{m_{\phi}^{4}}
\left[
(Y_{e\tau} Y_{\mu\mu})^2 
+(Y_{\mu\tau} Y_{e \mu})^2
+\fr12  Y_{e\tau} Y_{\mu\mu}Y_{\mu\tau} Y_{e \mu}
\right],
\nn
Br(\tau^{-}\to e^{+}\mu^{-}\mu^{-})
&=\fr53  \cdot \fr{\tau_{\tau}}{ 2^{12} \pi^{3}}\fr{m_{\tau}^{5}}{m_{\phi}^{4}}
 (Y_{\mu \tau}Y_{e\mu})^2,
\nn
Br(\tau^{-}\to  \mu^{-}\mu^{+}\mu^{-})
&=\fr53  \cdot \fr{\tau_{\tau}}{ 2^{11} \pi^{3}}\fr{m_{\tau}^{5}}{m_{\phi}^{4}}
 (Y_{\mu \tau}Y_{\mu\mu})^2,
\end{align}
where $\tau_{\mu(\tau)}$ is the life time of $\mu(\tau)$ lepton.
From those expressions, it is easy to see that the ratios
of the branching ratios    only depends on the 
texture of the Yukawa coupling matrix if the approximate relation of Eq.(\ref{6zeroSimple})
and Eq.({\ref{4zeroSimple}}) are used. For example, in the four texture-zero base ansatz
the predicted ratio $R$ between $\mu\to 3e$ and $\tau\to 3e$ is  
\begin{align}
R\equiv\fr{\mathcal{B}r(\mu\to e^{-}e^{+}e^{-})}{\mathcal{B}r(\tau\to  e^{-}e^{+}e^{-})}
\approx \fr{\tau_{\mu}}{\tau_{\tau}}
\left(\fr{m_{\mu}} {m_{\tau}}\right)^5, 
\end{align}
while the one   in the six texture-zero based Cheng-Sher ansatz (denoted by $R'$) is 
\begin{align}
R' \approx\fr{\tau_{\mu}}{\tau_{\tau}}
\left(\fr{m_{\mu}} {m_{\tau}}\right)^6.  
\end{align}
Thus, in the four texture-zero based texture  the ratio is larger by a factor of $m_{\tau}/m_{\mu}$. 
 
Using the allowed range of $\xi^{e}$ and $\xi^{e'}$ from Eq.(\ref{constraint}), the 
predicted branching ratios are obtained and summarized in  table.\ref{tab1}. 
\begin{table}[htb]\label{tab1}
\caption{Predicted branching ratios  for the three body lepton number violation decay modes in two ansatz. }
\begin{center}
\begin{ruledtabular}
\begin{tabular}{cccc}
decay mode & four texture-zero based ansatz  & Cheng-Sher ansatz  & Exp. bound\cite{Hagiwara:2002fs}\\\hline
$\mu^{-}\to e^{-}e^{+}e^{-}   $            &\sci{7.49}{-17} &\sci{2.31}{-19}&\sci{1.0}{-12}\\
$\tau^{-}\to e^{-}e^{+}e^{-}   $            &\sci{3.33}{-17} &\sci{1.85}{-18}&\sci{2.9}{-6}\\
$\tau^{-}\to \mu^{-}e^{+}e^{-} $          &\sci{6.67}{-15} &\sci{3.71}{-16}&\sci{1.7}{-6}\\
$\tau^{-}\to \mu^{+}e^{-}e^{-}$           &\sci{3.33}{-15} &\sci{1.85}{-16}&\sci{1.5}{-6}\\
$\tau^{-}\to e^{-}\mu^{+}\mu^{-} $      &\sci{1.33}{-12 }&\sci{7.41}{-14}&\sci{1.8}{-6}     \\ 
$\tau^{-}\to e^{+}\mu^{-}\mu^{-} $      &\sci{6.67}{-13} &\sci{3.71}{-14}&\sci{1.5}{-6}\\
$\tau^{-}\to \mu^{-}\mu^{+}\mu^{-} $   &\sci{1.33}{-10}&\sci{7.41}{-12}&\sci{1.9}{-6}           
\end{tabular}
\end{ruledtabular}
\end{center}
\end{table}
It is found that in all these decay modes the four texture-zero based ansatz predicts
larger branching ratios.  For example, in decay mode $\mu\to 3e$, their difference is
larger by two order of magnitudes. The four texture-zero base ansatz can give a
prediction of \sci{7.5}{-17} while the six texture-zero based ansatz predicts only
\sci{2.3}{-19}.  Similarly, in decay mode $\tau\to 3\mu$, the prediction from the four texture-zero
based ansatz is also much  larger, and can reach \sci{1.3}{-10}, while  the one from 
the Cheng-Sher ansatz is   \sci{7.4}{-12}.
Thus those decay modes, once discovered by the experiments,  will not only indicate the new
physics beyond the SM  but also   distinguish different models.

In summary, we have discussed the textures of the Yukawa coupling matrices in the general
2HDM.  In the framework of the parallel texture, the suppression of the flavor changing
neutral current (FCNC) can be achieved.  We have proposed a four texture-zero based ansatz
on the Yukawa coupling matrices which has several advantages over the six texture-zero
based ansatz proposed by Cheng and Sher.  It is found that this new ansatz is more
predictive in the lepton sector. As an example, the contribution from the general 2HDM to
the lepton number violation decay modes $\ell\to \ell_{1}\ell_{2}\ell_{3}$ are calculated
in both ansatz.  The results show that after considering the constraints from $\mu\to
e\gamma$, the four texture-zero based ansatz can predicts a decay rates two order of
magnitudes higher than the six texture-zero based one, and the branching ratio of $\mu\to
3e$ and $\tau\to 3\mu$ can reach \sci{7.5}{-17} and \sci{1.3}{-10} respectively.
\begin{acknowledgments}
The author is indebted to Y.L. Wu for helpful discussions. This work is
 support by the Alexander von Humboldt foundation.  
\end{acknowledgments}
\bibliographystyle{apsrev}
\bibliography{reflist}
\end{fmffile}\end{document}